\documentclass{ifacconf}

\usepackage{graphicx}      
\usepackage{natbib}        
\usepackage{amsmath,amssymb,amsfonts}
\usepackage{color}
\usepackage{url}

\newcommand{\x}{\mathbf{x}}
\newcommand{\y}{\mathbf{y}}
\newcommand{\z}{\mathbf{z}}
\newcommand{\p}{\mathbf{p}}
\newcommand{\ubold}{\mathbf{u}}
\newcommand{\A}{\mathcal{A}}
\newcommand{\M}{\mathcal{M}}

\newcommand{\Z}{\mathbb{Z}}
\newcommand{\V}{\mathcal{V}}
\newcommand{\E}{\mathcal{E}}
\newcommand{\DOA}{\mathrm{DOA}}
\newcommand{\post}{\mathrm{post}}
\newcommand{\fault}{\mathrm{fault}}
\newcommand{\pre}{\mathrm{pre}}
\newcommand{\fpost}{f^\mathrm{post}}
\newcommand{\ffault}{f^\mathrm{fault}}

\newcommand{\bdot}{\textbf{.}}
\newcommand{\safe}{\mathrm{safe}}
\newcommand{\unsafe}{\mathrm{unsafe}}
\newcommand{\C}{\mathrm{C}}
\newcommand{\F}{\mathrm{F}}

\newcommand{\mbbR}{\mathbb{R}}

\newcommand{\blk}{\color{black}}

\newtheorem{definition}{Definition}

\begin{document}
\begin{frontmatter}

\title{Critical Clearing Time Estimates of Power Grid Faults via a Set-Based Method\thanksref{footnoteinfo}} 

\thanks[footnoteinfo]{Stefan Streif thanks Tennet TSO GmbH for funding this research.}

\author[First]{Willem Esterhuizen}
\author[First]{Gyula Molnár}
\author[First]{Tim Aschenbruck}
\author[First]{Franz Rußwurm}
\author[Second]{Halil Askan}
\author[First]{Stefan Streif}

\address[First]{Automatic Control and System Dynamics, Technische Universität Chemnitz, Germany (e-mail: willem-daniel.esterhuizen@tu-ilmenau.de/gyula.molnar/tim.aschenbruck/franz.russwurm/stefan.streif@etit.tu-chemnitz.de).}
\address[Second]{TenneT TSO GmbH, Bayreuth, Germany (e-mail: halil.askan@tennet.eu).}

\begin{abstract}                
	This paper is concerned with estimating critical clearing times in the transient stability problem of power grids without extensive time-domain simulations. We consider a high-dimensional post-fault system (the grid after the fault is cleared) which we decouple into many smaller subsystems. Then, for each subsystem, we find the so-called safety sets and simulate the faulted system once to deduce the so-called safe and unsafe critical clearing times, which specify the intervals of time over which the fault may remain active before safety is compromised. We demonstrate the approach with a numerical example involving the IEEE 14 bus system.
\end{abstract}


\end{frontmatter}

\section{Introduction}

The transient stability problem is concerned with the behaviour of the rotor angles and rotor angle velocities of networked synchronous machines after a contingency occurs in the grid, such as a three-phase fault. The question is whether the machines will regain synchronism after the fault has been cleared (that is, after corrective action has been taken, such as the isolation of problematic buses), see \cite{kundur2004definition}, \cite{kundur2022power}. The \emph{critical clearing time} (CCT) or \emph{critical fault clearing time} (CFCT) is the longest amount of time over which the fault may remain uncleared. After this time the machines may not regain synchronism, or the system may settle at an undesired equilibrium point.



A number of methods exist in the literature to classify the post-fault state. These include approaches that use direct methods, see for example \cite{ribbens1985direct}, \cite{varaiya1985direct} and \cite{vu2015lyapunov}; time-domain simulations, as in \cite{chan2002time, zarate2009securing,chiang2010line, huang2012largeScale_TIME_DOM_SIM, nagel2013highPerfomance_Time_domain_sim}; approaches that combine both of these, see \cite{kyesswa2019hybrid}; and set-based methods, see \cite{althoff2012transient} \cite{jin2010reachability}, and \cite{oustry2019maximal}. However, the most reliable way to find a fault's critical clearing time is to consider a detailed model of the grid and to do extensive time-domain simulations. Typically, the fault system is simulated for some duration, say $t_1$ seconds. Then, a switch is made to the post-fault system and one checks if the machines eventually synchronise after further simulation. If this is true, a longer duration is chosen, say $t_2 > t_1$, and the simulation is done again. One iterates in this way until a maximal time duration is found, which is then the fault's critical clearing time. 

In this paper we propose a new set-based method to estimate critical clearing times that does not rely on extensive time-domain simulations. Following the ideas introduced in 
\cite{aschenbruck2020transient}, we decompose the high-dimensional post-fault system into many smaller systems and find their so-called \emph{safety sets}. We then simulate the fault-on system only once to obtain \emph{safe} and \emph{unsafe} critical clearing times (new notions we define in the paper). The idea is that practitioners might find the safety sets for many post-fault systems (offline) and then deduce the safe/unsafe CCTs from the single simulation of the fault-on system, when this information becomes available.

The paper's outline is as follows. In Section~\ref{sec:model} we cover the details of how we model power grid faults. Section~\ref{sec:approach} presents the new set-based approach for estimating critical clearing times, and Section~\ref{sec:numerics} presents a detailed example of a fault on the standard IEEE 14-bus system. In Section~\ref{sec:discussion} we have a discussion of the new approach's pros and cons, and point out directions for future research. 

\subsection*{Notation}

Consider a graph with $m\in\mathbb{N}$ nodes, where $\mathbb{N} = \{1,2\dots\}$ denotes the natural numbers. For node index $i\in\{1,2,\dots,m\}$, let $\mathcal{J}(i):= \{j_{i1}, j_{i2}, \dots, j_{in_i}\}$ refer to $i$'s neighbouring node indices, where $n_i := \#\mathcal{J}(i)$. Thus, $j_{ik}$, with $k\in\{1,2,\dots,n_i\}$, is node $i$'s $k$-th neighbour. $\mathbb{S}$ denotes the unit circle and $\mathbb{T}^m = \mathbb{S}\times\dots\times\mathbb{S}$ the $m$ dimensional torus, that is, the product of $m$ circles. $\mathbf{1}_{m}\in\mbbR^m$ (resp. $\mathbf{0}_m\in\mbbR^m$) denotes a column vector of $1$s (resp. $0$s). Given a matrix $M\in\mbbR^{q\times p}$, $M_{ij}$, $i\in\{1,\dots,q\}$ $j\in\{1,\dots,p\}$, denotes the entry in the $i$-th row and $j$-th column. Given a finite set of real numbers $S = \{x_k\}_{k\in\mathcal{I}}$, where $\mathcal{I}\subset\mathbb{N}$ is a finite index set, $M = \mathrm{diag}(S)\in\mbbR^{k\times k}$ is the diagonal matrix with $M_{ij} = x_i$ for $i=j$, and $M_{ij} = 0$ for $i\neq j$. 

\section{Modelling power grid faults}\label{sec:model}

We consider the so-called ``effective network'' model of connected oscillators, as presented in \cite[Ch. 2]{anderson2008power}. In this model each generator is represented by a voltage source behind a transient reactance with constant magnitude and time-varying angle (the ``classical'' generator model), the latter assumed to equal the generator's rotor angle. Mechanically, a generator is modelled as a mass (the rotor) that is driven by a mechanical torque, the equations of motion described by the \emph{swing equation}. Furthermore, it is assumed that all loads in the network are constant impedances and that the transmission lines are lossless (and thus purely imaginary). These simplifying assumptions do not result in large inaccuracies of the predicted rotor angle behaviours because of the short time intervals that are considered in transient stability studies (roughly one second, up to the ``first swing''). We will not present the full modelling details but rather refer the reader to the paper by \cite{nishikawa2015comparative}.

\sloppy

Given a network of $m\in\mathbb{N}$ generators, each with inertia constant $H_i\in\mbbR_{\geq 0}$, damping coefficient $D_i\in\mbbR_{\geq 0}$ and transient reactance $r_i\in\mbbR_{\geq 0}$, $i=1,2\dots,m$, the effective network model consists of a weighted $m$-vertex undirected graph $\mathcal{G} = (\V, \E)$, where $\V = \{1,2,\dots,m\}$ is the set of vertices; $\E= \{(i,j)\}$, $i,j\in \V$ is the set of edges; and $K_{ij}\in\mbbR_{\geq 0}$ is the edge weight between vertex $i$ and $j$. (Note that in the EN model all loads are lumped into the edges.) We let the square symmetric matrix $K\in\mbbR^{m\times m}$ denote the weight matrix. Each vertex is associated with a two-dimensional ordinary differential equation, with state $\x_{i}:=(x_{i1},x_{i2})^\top\in\mathbb{S}\times\mbbR$, that models an individual generator. Here, $x_{i1}$ denotes the $i$-th machine's rotor angle, whereas $x_{i2}$ denotes its rotor angle velocity. Let $\x_{\bullet 1} := (x_{11}, x_{21},\dots x_{m1})^\top\in\mathbb{T}^m$ denote all rotor angles; $\x_{\bullet 2} := (x_{12}, x_{22},\dots x_{m2})^\top\in\mathbb{R}^m$ denote all rotor angle velocities; and $\x := (\x_{\bullet 1}^\top, \x_{\bullet 2}^\top)^\top\in\mathbb{T}^{m}\times\mathbb{R}^{m}$ denote the full state.
\fussy For every $i\in\V$ the equations of motion read,
\begin{align}
	\dot{x}_{i1}(t) & = x_{i2}(t),\label{eq:EN_1}\\
	\dot{x}_{i2}(t) & = p_i - \sum_{j=1}^{m}K_{ij}\sin(x_{i1}(t) - x_{j1}(t)) - d_i x_{i2}(t),\label{eq:EN_2}
\end{align}
$t\in(-\infty,\infty)$, where the constants $p_i$, $d_i$ and $K_{ij}$ are determined from the generator parameters $H_i$, $D_i$ and $r_i$, the network's complex admittance matrix, as well as the system's power flow, see \cite{nishikawa2015comparative} for full details. Note that each generator is affected by the rotor angles of its neighbours.

As in \cite[Ch. 2]{anderson2008power} and \cite{varaiya1985direct}, we model a fault on the grid as a switch between three distinct dynamical systems,
\begin{align*}
	\dot{\x}(t)& = f^{\pre}(\x(t)),& &t\in(-\infty, t_F),\\
	\dot{\x}(t)& = f^{\fault}(\x(t)),\, \x(t^F) = \hat{\x}^{\pre},&  &t\in [t^F, t^C),\\
	\dot{\x}(t)& = f^{\post}(\x(t)), \,\x(t^C) = \x^C,&  &t\in [t^C,\infty),
\end{align*}
where $f^{\pre}(\bdot)$, $f^{\fault}(\bdot)$ and $f^{\post}(\bdot)$ (all mapping $\mathbb{T}^{m}\times\mathbb{R}^{m}$ to $\mathbb{T}^{m}\times\mathbb{R}^{m}$) denote the \emph{pre-fault}, \emph{fault-on} and \emph{post-fault} dynamics, respectively. We let $\hat{\x}^{\pre} \in\mathbb{T}^{m}\times\mathbb{R}^{m}$ and $\hat{\x}^{\post} \in\mathbb{T}^{m}\times\mathbb{R}^{m}$ denote the \emph{pre-fault equilibrium} and \emph{post-fault equilibrium}, respectively (that is $f^{\pre}(\hat{\x}^{\pre}) = f^{\post}(\hat{\x}^{\post}) = \mathbf{0}$). Thus, we assume that the system was at steady state before the fault occurred. The times $t^F\in\mathbb{R}$ and $t^C\in\mathbb{R}$,  $t^F\leq t^C$ are the \emph{fault time} and \emph{fault-clearing time}, respectively. Throughout the paper we assume that the dynamics $f^{\pre}$, $f^{\fault}$ and $f^{\post}$ are of the form \eqref{eq:EN_1}-\eqref{eq:EN_2}, each with a (possibly) different number of vertices and edges, and (possibly) different constants $(p_i^{\pre}$, $d_i^{\pre}$, $K_{ij}^{\pre})$, $(p_i^{\fault}$, $d_i^{\fault}$, $K_{ij}^{\fault})$ and $(p_i^{\post}$, $d_i^{\post}$, $K_{ij}^{\post}$). The reader may verify that due to the regularity of these equations, for every initial condition $\x^0\in \mathbb{T}^{m}\times\mathbb{R}^{m}$ there exists a unique solution to \eqref{eq:EN_1}-\eqref{eq:EN_2} defined for all $t\in(-\infty,\infty)$. With $\x(t_0) = \x^0$, $t_0\in\mbbR$, we let $\x^{\pre}(t; \x^0)$ (resp. $\x^{\fault}(t; \x^0)$ and $\x^{\post}(t; \x^0)$) denote the solution to the pre-fault (resp. fault-on and post-fault) system at time $t\in[t_0,\infty)$. The state $\x^C := \x^{\fault}(t^C; \hat{\x}^{\pre})$ is called the \emph{clearing state}.

\section{Set-based approach to estimate critical clearing times}\label{sec:approach}

Before we present the details of our approach in Subsection~\ref{subsec:approach} we cover a few preliminaries.

%

\subsection{Network synchronization and stability}

We quote a few concepts and results from the paper by \cite{dorfler2013synchronization}. Consider the system \eqref{eq:EN_1}-\eqref{eq:EN_2} and define
\[
	\Delta(\gamma):=\{\x_{\bullet 1}\in\mathbb{T}^m : |x_{i1} - x_{j1}| \leq \gamma,\,\,\forall (i,j)\in \mathcal{E}\},
\]
where $\gamma\in\mathbb{S}$. If $\x_{\bullet 1}\in\Delta(\gamma)$ we say that the angles are \emph{cohesive with angle $\gamma$}. A solution $\x(t;\x^0):\mbbR_{\geq 0}\rightarrow \mathbb{T}^m\times\mathbb{R}^m$ to the couple oscillator model, \eqref{eq:EN_1} - \eqref{eq:EN_2}, initiating at time $t=0$ is said to be \emph{synchronised} if there exists a $\gamma\in\mathbb{S}$ and an $\omega_{\mathrm{synch}}\in\mathbb{R}$ such that $\x_{\bullet 1}(0;\x^0)\in\Delta(\gamma)$, $\x_{\bullet 1}(t;\x^0) = \x_{\bullet 1}(0;\x^0) + \omega_{\mathrm{synch}} \mathbf{1}_{m}t$ (mod 2$\pi$), and $\x_{\bullet 2}(t;\x^0) = \omega_{\mathrm{synch}} \mathbf{1}_{m}$ for all $t\geq 0$. Thus, the system \eqref{eq:EN_1} - \eqref{eq:EN_2} is synchronised at time $t=0$ with angle $\gamma$ if all machines are rotating at the same angular velocity and every rotor angle is within $\gamma$ of its neighbouring angles, for all future time. This synchronous frequency is explicitly given by $\omega_{\mathrm{synch}} = \sum_i p_i / \sum_i d_i$ (see the supporting information of \cite{dorfler2013synchronization}), and so after redefining the state variables relative to this value we may, without loss of generality, assume that equilibrium points (if they exist) are located on the $x_1$ axis. 

The following proposition presents easily checkable sufficient conditions that imply the existence of a unique stable equilibrium.
\begin{prop}[\cite{dorfler2013synchronization}]
	For a given $\gamma\in\mathbb{S}$, the coupled oscillator model \eqref{eq:EN_1} - \eqref{eq:EN_2} has a unique and stable equilibrium point $\hat{\x} := (\hat{\x}_{\bullet 1}^\top, \mathbf{0}_m^\top)^\top\in \mathbb{T}^{m}\times\mathbb{R}^m$ (mod 2$\pi$) with $\hat{\x}_{\bullet 1}\in\Delta(\gamma)$ if:
	\begin{equation}
		\Vert L^{\dagger}\p\Vert_{\mathcal{E},\infty} \leq \sin(\gamma).\label{eq:stab_test}
	\end{equation}
\end{prop}
Here, $L^\dagger\in\mbbR^{m\times m}$ is the Moore-Penrose inverse of the graph's Laplacian,
\[
	L := \mathrm{diag}\left( \{ \sum_{j = 1}^{n} K_{ij} \}_{i=1,\dots,m} \right) - K.
\]
Recall that $K$ is the network's weight matrix. For some $\y\in\mbbR^m$, $\Vert \y  \Vert_{\mathcal{E},\infty} := \max_{(i,j)\in\E} |y_i - y_j|$ denotes the \emph{worst-case dissimilarity} over the edges $\E$, and $\p:=(p_1,p_2,\dots,p_m)^\top$ is the vector of all power injections. Note that even if there exists a stable post-fault equilibrium it is not guaranteed that the state will settle there from an arbitrary initial condition $\x^0$. 

\subsection{The new set-based approach}\label{subsec:approach}

Ideally, we would like find the \emph{domain of attraction}, see for example \cite[Ch. 4]{khalil2002nonlinear}, of the post-fault system's equilibrium point,
\[
\DOA(\hat{\x}^{\post}) := \{\x^0\in\mathbb{T}^m\times\mathbb{R}^{m} :  \lim_{t\rightarrow\infty}\x^{\post}(t;\x^0) = \hat{\x}^{\post}\},
\]
and then simulate $\ffault$ from the pre-fault equilibrium $\hat{\x}^{\pre}$ until the state intersects the boundary of $\DOA(\hat{\x}^{\post})$. We could then find the time,
\[
t^{\mathrm{CCT}} := \sup_{t\geq 0} \{|t - t^F| : \x^{\fault}(t;\hat{\x}^{\pre}) \in \DOA^{\post}(\hat{\x}^{\post}) \},
\]
and report $t^{\mathrm{CCT}}$ as the longest amount of time for which the fault may be active. However, it is very difficult, if not impossible, to find the DOA for systems with as high dimension as those in power systems.

The idea we now present is to \emph{decouple} the $2m$-dimensional system into $m$ two-dimensional systems. Then, we find each individual generator's \emph{safety sets}, and simulate $\ffault$ to obtain safe and unsafe critical clearing times (which we define presently). The computational effort of finding the safety sets for the decoupled systems now scales linearly with the dimension.

\subsubsection{Step 1.}

We gather the information required for the set-based approach. This comprises the pre-fault equilibrium, $\hat{\x}^{\pre}:= (\hat{x}_{11}^{\pre}, \hat{x}_{21}^{\pre},\dots, \hat{x}_{m1}^{\pre}, \mathbf{0}_m^\top)$, the fault-on system, $\ffault$, and the post-fault system $\fpost$.

\subsubsection{Step 2.}

We consider $\fpost$ and check condition \eqref{eq:stab_test} to see if there exists a stable equilibrium point, $\hat{\x}^{\post}:= (\hat{x}_{11}^{\post}, \hat{x}_{21}^{\post},\dots, \hat{x}_{m1}^{\post}, \mathbf{0}_m^\top)$. If the test passes we find $\hat{\x}^{\post}$ and impose constraints about it and $\hat{\x}^{\pre}$ as follows: for each $i\in\{1,2,\dots,m\}$ choose pairs $(\underline{x}_{i1}, \overline{x}_{i1})$ such that
\begin{align*}
	\underline{x}_{i1}&\leq \hat{x}^{\pre}_{i1}\leq  \overline{x}_{i1},\\
	\underline{x}_{i1}&\leq \hat{x}^{\post}_{i1}\leq  \overline{x}_{i1}.
\end{align*}

\subsubsection{Step 3.}

\sloppy
We now decouple the system into $m$ two-dimensional systems subject to state and input constraints. Let $\z_i\in\mathbb{S}\times\mbbR$ denote the state of the $i$-th decoupled post-fault system; let 
\[
	\mathbb{Z}_i:= [\underline{x}_{i1}, \overline{x}_{i1}]\times\mathbb{R},
\]
\fussy denote the $i$-th system's local state constraints. Now, considering the neighbours of machine $i$ define,
\begin{align*}
	\ubold_i := (u_{i1}, u_{i2}, \dots, u_{in_i})^\top\in\mathbb{R}^{n_i},
\end{align*}
and let
\[
	\mathbb{U}_i:= [\underline{x}_{j_{i1}1}, \overline{x}_{j_{i1}1}]\times [\underline{x}_{j_{i2}1}, \overline{x}_{j_{i2}1}] \times \dots \times [\underline{x}_{j_{in_i}1}, \overline{x}_{j_{in_i}1}].
\]
Recall that $n_i$ is the number of neighbours of vertex $i$. We assume that $\ubold_i$ is a piece-wise continuous function (a local ``input'')  mapping $[0,\infty)\rightarrow \mathbb{U}_i$. The decoupled post fault systems now read, for each $i\in\{1,2,\dots, m\}$,
\begin{align}
	\dot{z}_{i1}(t) & = z_{i2}(t),\label{eq:decoupled_1}\\
	\dot{z}_{i2}(t) & = p_i^{\post} - \sum_{j = 1}^{n_i}K_{ij}^{\post}\sin(z_{i1}(t) - u_{ij}(t)) - d_i^{\post}z_{i2}(t),\label{eq:decoupled_2}
\end{align}
with state and input constraints,
\begin{align*}
	\z_{i}(t) & \in \mathbb{Z}_i\subset\mathbb{S}\times\mathbb{R},\\
	\ubold_i(t) & \in \mathbb{U}_i\subset\mathbb{R}^{n_i},
\end{align*}
$t\in[0,\infty)$. Note that the constants $p_i^{\post}$, $K_{ij}^{\post}$ and $d_i^{\post}$ are those of the post-fault system.

\subsubsection{Step 4.}

For each decoupled machine we find two subsets of its two-dimensional state space: the \emph{admissible set}, see \cite{de2013barriers},
\[
	\mathcal{A}_i := \{\z_i^0\in\mathbb{Z}_i: \exists \ubold_i \in \mathcal{U}_i,\ \z_i(t;\z_i^0,\ubold_i)\in\mathbb{Z}_i\ \forall t\in[0,\infty)\},
\]
and the \emph{maximal robust positively invariant set} (MRPI) contained in $\mathbb{Z}_i$, see \cite{esterhuizen2020maximal},
\[
\mathcal{M}_i := \{\z_i^0\in\mathbb{Z}_i: \z_i(t;\z_i^0,\ubold_i)\in\mathbb{Z}_i\ \forall t\in[0,\infty), \forall \ubold_i\in\mathcal{U}_i\}.
\]
Here, $\mathcal{U}_i$ is the space of piece-wise continuous functions mapping $[0,\infty)$ to $\mathbb{U}_i$, and $\z_i(t;\z_i^0,\ubold_i)$ is the solution to \eqref{eq:decoupled_1}-\eqref{eq:decoupled_2} at time $t\geq 0$ from initial condition $\z_i^0\in\mathbb{S}\times\mbbR$ with input $\ubold_i\in\mathcal{U}_i$.

\subsubsection{Step 5.}

We simulate the fault-on dynamics $f^{\fault}$ from the pre-fault equilibrium, $\hat{\x}^{\pre}$, and define, for each machine, the following two quantities:
\[
t_{\M_i}:=\inf\{t\in\mathbb{R}_{\geq 0}:\x_{i}^{\mathrm{\fault}}(t;\hat{\x}^{\pre})\in\partial\M_i\},
\]
and,
\[
t_{\A_i}:=\inf\{t\in\mathbb{R}_{\geq 0}:\x_{i}^{\mathrm{\fault}}(t;\hat{\x}^{\pre})\in\partial\A_i\}.
\]
We then define the \emph{safe CCT} and \emph{unsafe CCT} as follows,
\[
t^{\safe}:=\inf_{i\in\{1,2,\dots,M\}}\{t_{\M_i} \}
\]
and
\[
t^{\unsafe}:=\inf_{i\in\{1,2,\dots,M\}}\{t_{\A_i} \},
\]
respectively.

Before presenting the last step we introduce the following definitions.
\begin{definition}
	The post-fault system $f^{\post}$ is said to be \emph{safe} provided that $\x^{\post}(t;\x^C)\in\Z$, for all $t\in[t^C,\infty)$; and \emph{unsafe} if it is not safe.
\end{definition}

\subsubsection{Step 6.}

We now deduce the criticality of the fault from the following proposition, which is a slight modification of a result first reported in \cite{aschenbruck2020transient}.
\begin{prop}\label{prop_main}
	Consider the solution of the fault-on dynamics initiating from the pre-fault equilibrium, $\x^{\fault}(t;\hat{\x}^{\pre})$, for $t\in[t^{\F},t^{\C}]$.
	\begin{itemize}
		\item If $t^C \leq t^{\safe}$ then $f^{\post}$ is \emph{safe}.
		\item If $t^C \geq t^{\unsafe}$ then $f^{\post}$ is \emph{unsafe}.
		\item If $t^C \in (t^{\safe}, t^{\unsafe})$ then $f^{\post}$ is \emph{potentially safe}.
	\end{itemize}
	By potentially safe, we mean that $f^{\post}$ is either safe or unsafe.
\end{prop}
To aid in the proof of this proposition we introduce some more notation. Let $\ubold := \Pi_{i=1}^{m} \ubold_i$, $\mathcal{U} := \Pi_{i=1}^{m} \mathcal{U}_i$, $\M:= \Pi_{i=1}^{m} \M_i$ and $\mathbb{Z} := \Pi_{i=1}^{m}\mathbb{Z}_i$. Furthermore, let $\mathcal{U}^T$ denote all piece-wise continuous functions mapping $[0,T]$ to $\Pi_{i=1}^{m}\mathbb{U}_i$, and let $\A^{\mathsf{C}}$ denote the complement of $\A$.

\begin{pf}[Prop. 2]
	Consider an aribitrary initial state $\x^0\in\Z$ and an arbitrary interval of time $[0,T]$, $T< \infty$. Then, we see that $\x^{\post}(t;\x^0)\in\Z$ for all $t\in[0, T]$ if and only if there exists a $\ubold\in\mathcal{U}^T$ such that $\z(t;\x^0,\ubold) = \x(t;\x^0)$ for all $t\in[0,T]$. Now, if $t^C\leq t^{\safe}$ then $\x^C\in\M$, which implies that $\z(t;\x^C,\ubold)\in\Z$ for all $t\in[t^C,\infty)$, for all $\ubold\in\mathcal{U}$. Therefore, there exists a $\ubold\in\mathcal{U}$ such that $\z(t;\x^C,\ubold) = \x^{\post}(t;\x^C)$ for all $t\in[t^C,\infty)$, which implies that $\x^{\post}(t;\x^C)\in\Z$ for all $t\in[t^C,\infty)$, and so $f^{\post}$ is safe. Similary, if $t^C\geq t^{\unsafe}$, then $\x^C\in\A^{\mathsf{C}}$. Thus, for all $\ubold\in\mathcal{U}$ there exists a $t(\ubold)<\infty$ such that $\z(t(\ubold);x^C,\ubold) \notin \Z$, which implies that there exists a $\hat{t}<\infty$ such that $\x(\hat t;x^C)\notin \Z$. Thus $f^{\post}$ is unsafe. Finally, if $t^C\in(t^{\safe}, t^{\unsafe})$ then $\x^C\in \A$ and there exists a $\ubold\in\mathcal{U}$ such that $\z(t;\x^C, \ubold) \in \Z$ for all future time, but this realisation of the decoupled system does not necessarily correspond to the true solution of the post-fault system from $\x^C$. Thus, this case is indeterminate, and $f^{\post}$ is potentially safe.
	

	\blk
%
%
%
\end{pf}

\subsection{An optimization problem}
\sloppy

Let $\underline{\x} := (\underline{x}_{11}, \underline{x}_{21},\dots,\underline{x}_{M1})^\top\in\mathbb{R}^{M}$ and $\overline{\x} := (\overline{x}_{11}, \overline{x}_{21},\dots,\overline{x}_{M1})^\top\in\mathbb{R}^{M}$. For given values of $\underline{\x}$ and $\overline{\x}$ the safety sets $\A_i$ and $\M_i$ may have small volume, or even be empty, which would result in small values for $t_{\safe}$ and $t_{\unsafe}$. Thus, we would like to choose these bounds to maximise the set volumes and we can consider the following optimization problem,
\begin{align*}
	\max_{\overline{\x}, \underline{\x}} & \,\, J(\overline{\x}, \underline{\x}),\\
	\mathrm{s.t.}&\,\, \underline{x}_{i1}\leq \overline{x}^{\pre}_{i1}\leq  \overline{x}_{i1},\,\,i\in\{1,2,\dots,M\},\\
	&\,\, \underline{x}_{i1}\leq \overline{x}^{\post}_{i1}\leq  \overline{x}_{i1},\,\,i\in\{1,2,\dots,M\},
\end{align*}
where $J(\overline{\x}, \underline{\x}) = \sum_i^{M} \mathrm{V}(\M_i)$, and $\mathrm{V}(\M_i)$ is the volume of the $i$-th MRPI. Unfortunately, this is not a straight-forward optimization problem to address, as the set volumes cannot conveniently be expressed as a function of the bounds $\overline{\x}$ and $\underline{\x}$.

\fussy
However, the paper \cite{aschenbruck2020transient} conducted an in-depth study of the safety sets for the system \eqref{eq:EN_1}-\eqref{eq:EN_2}. Using the \emph{theory of barriers}, see \cite{de2013barriers} and \cite{esterhuizen2020maximal}, the paper showed how the exact safety sets  $\A_i$ and $\M_i$ (which may be empty) can easily be obtained for a given vector $(\overline{\x}, \underline{\x})$ simply through the integration of the system equations. In the next section we attempt to find ``good''  values  $(\overline{\x}, \underline{\x})$ through black-box optimization.

\section{Numerics}\label{sec:numerics}
In the following we will consider the IEEE 14-bus test system, see \url{case14} of the Matpower manual, \cite{zimmermanMatpowerManual}, to provide a numerical example for set-based CCT computation.

\subsection{IEEE 14-bus Test System}
\begin{figure}[hbt!] 
	\begin{center}
		\includegraphics[width=8.4cm]{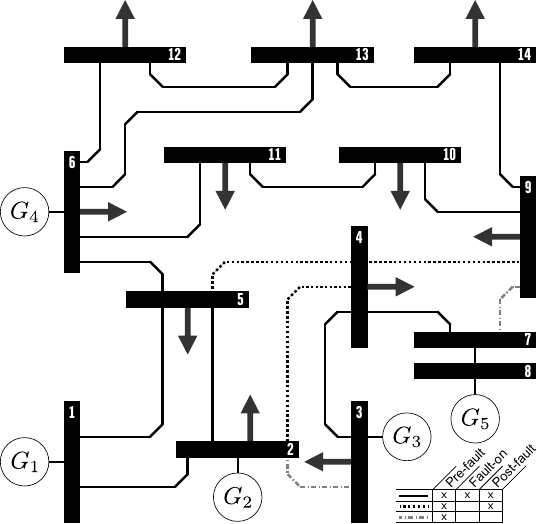}    
		\caption{The IEEE 14-bus Test System.} 
		\label{fig:14bus}
	\end{center}
\end{figure}

Fig. \ref{fig:14bus} shows the system's bus diagram with fault-on and post-fault alterations marked. We assume that a contingency causes branches 2-3, 2-4, 4-5, 4-9, and 7-9 to temporarily fail and that after error clearance all branches but 2-3, and 7-9 are successfully reattached to the rest of the grid. This disconnection of the graph's edges could, for example, model three-phase faults that occur at either of the bus terminals between which a tranmission line runs. Table \ref{tb:dyn} summarizes the dynamic parameters considered. We use the code from \cite{nishikawa2015comparative}, along with the power flow solver from Matpower, \cite{zimmermanMatpower}, to obtain the three EN models (one for pre-fault, fault-on and post-fault, using the power flow of the pre-fault system).

Having defined system parameters, we determine the pre-fault and post-fault equilibrium states, $\hat{\x}^{\pre}$ and $\hat{\x}^{\post}$ respectively, then obtain bounds $\overline{x}_{i1}$ and $\underline{x}_{i1}$ via a \emph{black-box} optimization approach from \cite{molnar2021}, see Table \ref{tb:angularConstraints}. 
Finally, the fault-on trajectory is obtained through numerically forwards integrating the full ten-dimensional fault-on dynamics, $f^{\fault}$ from the pre-fault equilibrium point.

\begin{figure}[hbt!] 
	\begin{center}
		\includegraphics[width=8.4cm]{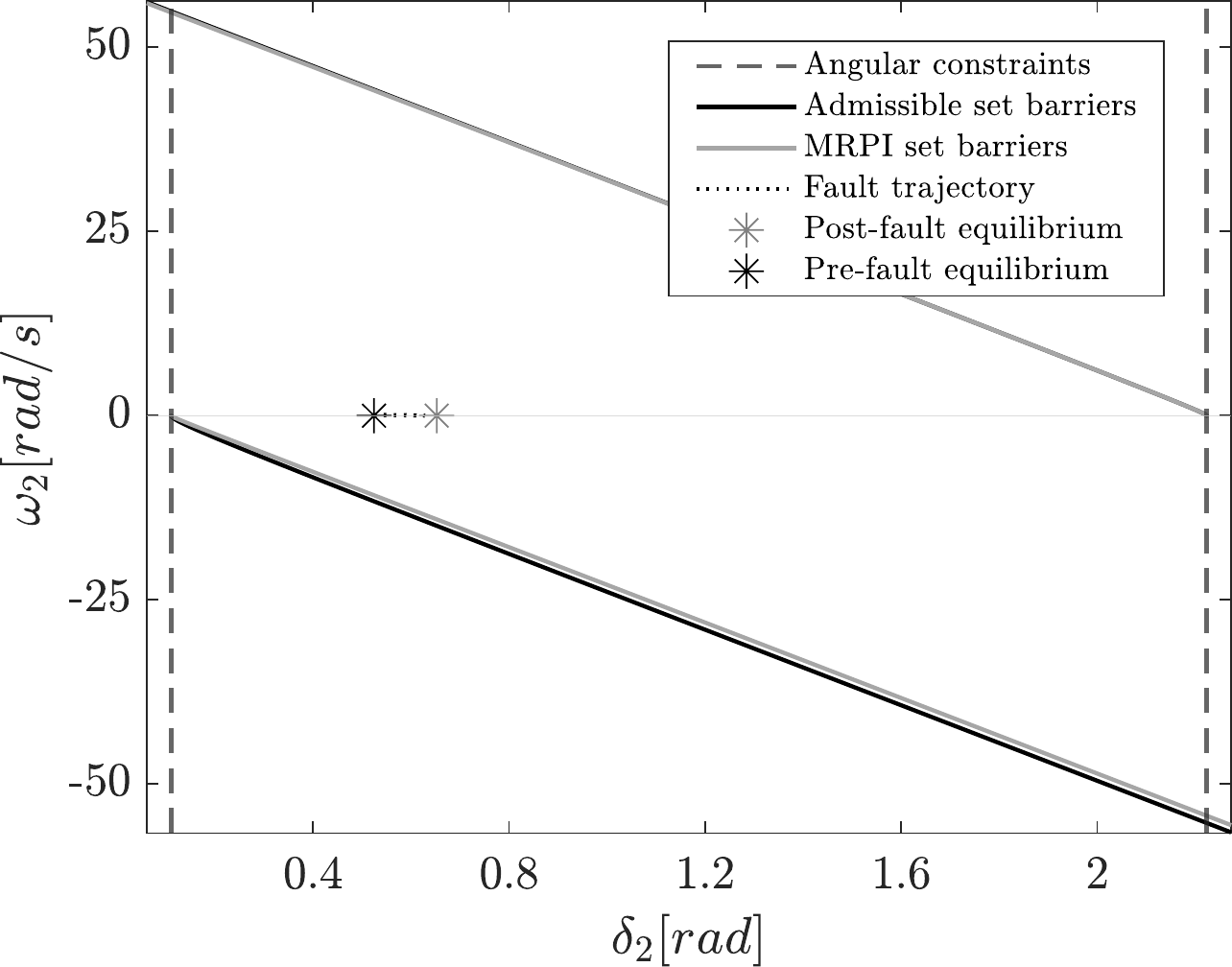}    
		\caption{Safety sets of generator 2, $G_2$, along with the fault-on trajectory.} 
		\label{fig:G2}
	\end{center}
\end{figure}

With the obtained bounds $\overline{x}_{i1}, \underline{x}_{i1}$ the MRPI of generator 1 is empty, $\M_1 = \varnothing$, though it has a nonempty admissible set $\mathcal{A}_1$. The phase diagrams of generators $G_3, G_4$ and $G_5$ look topologically analogous to that of $G_2$ (Fig. \ref{fig:G2}) with $\varnothing \neq \M_i \subset \A_i$ and $\{\hat{\x}^{\pre}_{i}, \x_i^{\mathrm{\fault}}(t), \hat{\x}^{\post}_{i}\} \subseteq \M_i$ for all $t$, for $i \in \{2 \ldots 5\}$. We furthermore see that the fault trajectory intersects the boundary of $\A_1$ after $t_{\A_1} \approx 1.3761 \, s$. Since $\M_1 = \varnothing$ we deduce that $t^{\safe} = 0$, but that $t^{\unsafe}=\inf_{i\in\{1,2,\dots,M\}}\{t_{\A_i} \} = t_{\A_1} \approx 1.3761 \, s$. That is, in this specific scenario we may pronounce that $f^{\fault}$ is never safe, potentially safe for $t^{C} < 1.3761 \, s$ and unsafe for $t^C \geq 1.3761 \, s$. Finally, because $\M_1$ is empty, we can say that $G_1$ is the \emph{critical generator} and that this is where emergency resources should be concentrated in case this fault scenario occurs. This is counter-intuitive because $G_1$ is in fact located further away from the faulted lines than $G_2$, $G_3$ and $G_5$.



\begin{table}[hbt!]
	\begin{center}
		\caption{Dynamic parameters: transient reactance $r_i$, inertia constant $H_i$, damping coefficient $D_i$, real power output $P_i$, and reactive power output $Q_i$ of each generator. Power unit (p.u.) quantities are with respect to the \mbox{100 MVA} system base. The system's reference frequency is \mbox{60 Hz}.}\label{tb:dyn}
		\begin{tabular}{crrrrr}
			$G_i$ & $r_i [\mbox{p.u.}]$ & $H_i [\mbox{s}]$ & $D_i [\mbox{p.u.}]$ & $P_i [\mbox{MW}]$ & $Q_i [\mbox{MVAr}]$  \\\hline
			$G_1$ & 0.0050 & 185.4630 & 0.5000 & 232.4 & -16.9  \\
			$G_2$ & 8.9916 & 12.9333  & 664.4750 & 40.0 & 42.4 \\ 
			$G_3$ & 16.9450 & 1.5781 & 989.5800 & 0 & 23.4 \\
			$G_4$ & 2.2604 & 0.0010 & 780.8900 & 0 & 12.2 \\
			$G_5$ & 20.0000 & 0.4621 & 772.3950 & 0 & 17.4 \\ \hline
		\end{tabular}
	\end{center}
\end{table}
\begin{table}[hbt!]
	\begin{center}
		\caption{Equilibrium states and optimal bound candidates.}\label{tb:angularConstraints}
		\begin{tabular}{crrr}
			$G_i$ & $\hat{\x}^{\pre}_{i1} [\mbox{rad}]$ & $\underline{x}_{i1} [\mbox{rad}]$ & $\overline{x}_{i1} [\mbox{rad}]$ \\\hline
			$G_1$ & 0       & -0.3430 & 0.3903 \\
			$G_2$ & 0.6526  & 0.1110  & 2.2234 \\
			$G_3$ & -0.4383 & -1.0731 & 1.1325 \\
			$G_4$ & -0.3409 & -1.1704 & 1.1290 \\
			$G_5$ & -0.2484 & -1.3188 & 0.7816 \\ \hline
		\end{tabular}
	\end{center}
\end{table}

\begin{figure}[hbt!] 
	\begin{center}
		\includegraphics[width=8.4cm]{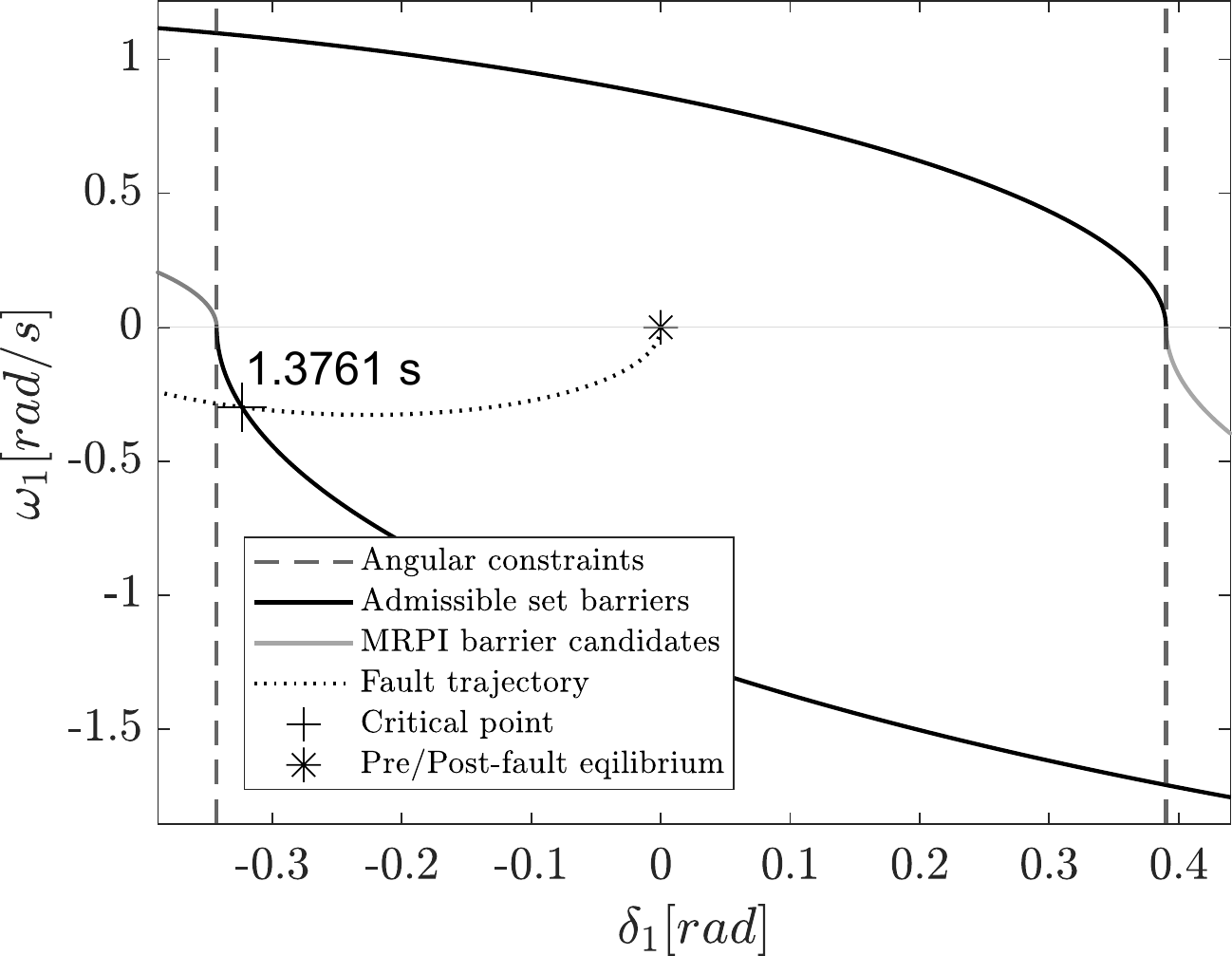}    
		\caption{Safety sets of generator 1, $G_1$, along with the fault-on trajectory.} 
		\label{fig:G1}
	\end{center}
\end{figure}

\section{Discussion and Perspectives}\label{sec:discussion}

As mentioned in the introduction, the idea is to find the safety sets for various post-fault grids offline and then to determine the safe/unsafe CCTs from a single simulation of the fault-on grid when this information becomes available. A downside of the approach is that, due to the decoupling of the system, there might not exist bounds $(\overline{\x}, \underline{\x})$ for which the sets $\M_i$ are nonempty. Thus, in finding the safety sets as opposed to the true DOA we drastically reduce the computational effort, but trade this for greater conservatism.

However, the results may still be useful even with conservatism in the predictions, especially in identifying highly stable/unstable faults. Thus, the approach could be used to ``pre-filter'' these faults, leaving the remaining ones to be investigated in more detail in simulation. The approach can also be used to identify particular problematic nodes associated with a fault (those with small or empty safety sets) as was shown in the numerical example. 

The main contrast between the safe/unsafe CCTs we introduce and traditional CCT is that we do not make a prediction regarding asymptotic stability: Proposition~\ref{prop_main} makes a statement about the state remaining within the set $\Z$. Extending the approach to also deduce asyptotic stability of the post-fault system could form the focus of future research. We note that every ``safe'' system we investigated settled at an equilibrium in simulation.

Future research should focus on applying the approach to much larger grids and comparing the obtained safe/unsafe CCTs to those calculated by standard commercial software, such as PowerFactory. Finally, although the optimization problem to obtain the best bounds $(\overline{\x}, \underline{\x})$ is not straightforward, it might be that black box methods (which are recieving a lot of recent interest, especially with respect to their application in machine learning) are effective at solving it. This also deserves more attention.

%
 

\bibliography{ifacconf}             

\end{document}